\documentclass[letterpaper,12pt]{article}
\usepackage[letterpaper,margin=1in]{geometry}
\usepackage{setspace}
\usepackage{amsmath}
\usepackage{amssymb}
\usepackage{mathrsfs}
\usepackage{bbm}
\usepackage{graphicx}
\usepackage{verbatim}
\usepackage{yfonts}
\usepackage{eufrak}
\usepackage{color}
\definecolor{darkblue}{cmyk}{0.9,0.9,0,0} 

\begin{document}
\thispagestyle{empty}
\begin{center}
\Large{\textbf{Higher Codimension Singularities Constructing Yang-Mills Tree Amplitudes}} \\
\vspace{10mm}
\large\text{Sayeh Rajabi}\footnote{srajabi@perimeterinstitute.ca} \\
\vspace{15mm}
\normalsize \textit{Perimeter Institute for Theoretical Physics,} \\
 \normalsize\textit{Waterloo, ON, Canada, N2L 2Y5} \\ \& \\
\normalsize\textit{Department of Physics and Astronomy $\&$ Guelph-Waterloo Physics Institute,} \\
\normalsize \textit{University of Waterloo, Waterloo, ON, Canada, N2L 3G1}\\ 
\end{center}
\vspace{20mm}
\abstract
\hspace{-6pt}Yang-Mills tree-level amplitudes contain singularities of codimension one like collinear and multi-particle factorizations, codimension two such as soft limits, as well as higher codimension singularities. Traditionally, BCFW-like deformations with one complex variable were used to explore collinear and multi-particle channels. Higher codimension singularities need more complex variables to be reached. In this paper, along with a discussion on higher singularities and the role of the global residue theorem in this analysis, we specifically consider soft singularities. This is done by extending Risager's deformation to a $\mathbb C^2$-plane, i.e., two complex variables. The two-complex-dimensional deformation is then used to recursively construct Yang-Mills tree amplitudes.
\newpage

\section{Introduction}
\setcounter{page}{1}
Since the introduction of the CSW expansion \cite{CSW} and BCFW recursion relations \cite{BCF, BCFW}, the methods have been extensively used in calculations of tree and loop-level amplitudes in QCD, $\mathcal{N}=4$ super Yang-Mills theory, general relativity and $\mathcal{N}=8$ supergravity. For a review of on-shell methods, see \cite{Bern review} and the references listed there, see also \cite{susy1, susy} for the supersymmetric extension of BCFW deformation. 

Inspired by these modern methods of S-matrix calculation and also twistor space formulation of BCFW, a dual formulation for the S-matrix of $\mathcal{N}=4$ SYM was proposed \cite{duality}.\footnote{The reader is referred to \cite{G1} for a sample of recent progress.} Using multi-dimensional complex analysis, this duality connects the leading singularities of planar $\text{N}^{k}$MHV amplitudes to simple contour integrals over the Grassmannian manifold of $k$-planes in $n$-dimensions. 

The idea of the BCFW technique is to apply the residue theorem to the complexified tree amplitude, $M(z)$, and construct the non-deformed amplitude recursively, 
\begin{eqnarray}
M(0)=\dfrac{1}{2\pi i}\oint_C dz\dfrac{M(z)}{z} -\displaystyle\sum_{z_{ij}}\text{Res}(\frac{M(z)}{z}),
\end{eqnarray}
where the contour of integration encloses all the poles of integrand, and $z_{ij}$ are poles of $M(z)$. 

In order to complexify the amplitude, two of the external momenta are shifted into complex plane preserving momentum conservation. In spinor-helicity notation, BCFW deformation is on the holomorphic and anti-holomorphic spinors of two different particles,
$$\lambda^k(z)=\lambda^k+z \lambda^l,\quad \tilde{\lambda}^l(z)=\tilde{\lambda}^l-z \tilde{\lambda}^k,$$ in such a way that $M(z)$ vanishes at infinity. In $\mathcal{N}=4$ super Yang-Mills, $\tilde{\eta}_l(z)=\tilde{\eta}_l-z\tilde{\eta}_k$, and any two particles can be deformed. Here we will only consider gluon amplitudes and the helicities of $k$ and $l$ can be anything except $(-,+)$. 

BCFW-like deformations were used to give a direct proof of the CSW-expansion of amplitudes in pure Yang-Mills by Risager \cite{Risager}, and in super Yang-Mills by Elvang et al \cite{elvang2}. Risager's deformation, applied only to $\tilde\lambda$ of negative helicity particles, contains an auxiliary anti-holomorphic spinor, $\tilde{\eta}$ \footnote{Not to be confused with the Grassmann parameter $\tilde{\eta}$ in the supersymmetric deformation.}, similar to CSW's reference spinor, 
$$\tilde{\lambda}^i\longrightarrow\tilde{\lambda}^i+z\alpha^i\tilde{\eta},$$ where $\alpha^i$ are constant. For an amplitude with $k$ negative helicities, we can fix two of $\alpha^i$'s using momentum conservation, $$\sum_i\alpha^i\lambda^i=0.$$Now in order to have BCFW-like recursion relations we must impose the condition under which the amplitude $M(z)$ vanishes while $z\rightarrow\infty$, which is precisely satisfied by this deformation. By an induction procedure, Risager showed that Yang-Mills tree amplitudes can be constructed using CSW rules.

On a $\mathbb C^2$-plane where $\tilde\lambda^i$ lives, Risager's deformation constrains the shifted spinor to a strip made by the original $\tilde\lambda^i$ and the shift, $z \alpha^i \tilde{\eta}$. The full $\mathbb C^2$-plane hence can not be reached by the shift. This observation suggests the idea of generalization of Risager's deformation in order for $\tilde\lambda^i$'s to have access to entire space. In fact it is more natural to deform spinors in a two-complex-dimensional plane, $\mathbb C^2$, for which we need two complex variables. The amplitude therefore depends on two complex variables and the generalization of the residue theorem to several variables, the global residue theorem (reviewed in section 3), can be applied. 

We follow the steps to reconstruct the physical non-deformed amplitude. The novelty is that now the amplitude generically receives contributions from channels that were not accessible before. In our examples in this paper these new contributions are soft limits (in each of which one of the external deformed momenta vanishes), and double-factorization channels. Therefore, the full amplitude can be reconstructed using two codimension one and a single codimension two singularities.

Using generalized deformations, we will have access to channels of interaction which are out of reach by one-variable BCFW or Risager's shifts. This is our main motivation for using complex multi-variable analysis in calculations of scattering amplitudes. In multi-variable analysis where all the shifts are linear, the Cauchy's theorem can be applied several times to build non-deformed amplitudes. However, with generic deformations (e.g. non-linear shifts), the only possible way to solve the multi-variable problem is applying the global residue theorem.  

Here we restrict ourselves to color-ordered Yang-Mills tree-level amplitudes. We discuss general BCFW-like deformations and the necessity of applying the global residue theorem in section 2. Risager's two-variable deformation is introduced in particular. In section 3, residues in multi-dimensional complex analysis and the global residue theorem, our mathematical tool, are briefly reviewed. Calculations of NMHV 5- and 6-particle amplitudes with two complex variables and appearance of soft terms, as new contributions, are given in sections 4 and 6 respectively. Section 5 generalizes the argument to the $n$-particle N$^{k-2}$MHV amplitudes. We discuss that with the introduced deformation, there is no more singular term in the corresponding residue theorem, except the known collinear, multi-particle and soft singularities. We finally make some concluding remarks in section 7.      

\section{General deformations and the global residue theorem}
Through general deformations on holomorphic and anti-holomorphic spinors, scattering amplitudes are generic functions of several complex variables. The simplest linear deformation with one variable is BCFW by which the Cauchy's theorem generates non-deformed amplitudes. In BCFW and also Risager's one-variable methods, not all but some of the singularities of amplitudes can be reached. These singularities are collinear, where two external momenta are orthogonal ($p^i.p^j=0$), and multi-particle ($(\sum_i p^i)^2=0$ for a subset of external particles) \footnote{One can think of a particular auxiliary spinor in Risager's deformation by which some external momenta can be soft, but generically soft singularities are not visible in this way. Consider the deformation $\tilde{\lambda}^i\rightarrow\tilde\lambda^i+z\alpha^i\tilde{\lambda}^1$ on negative helicity particles which include particle 1. It can be immediately seen that $\tilde{\lambda}^1(z)$ vanishes at $z^*=-1/\alpha^1$. This is in fact the pole of all the diagrams in which particle 1 is collinear with any other particle.}. We call them codimension one singularities where each of them can be determined by one condition on external momenta. It is clear that one complex variable in the deformation is enough for solving the condition and finding the pole.

Applying a linear two-variable Risager's deformation, amplitudes exhibit codimension two singularities: (codimension one)$\times$(codimension one), and soft singularities. The former corresponds to a two-factorization channel of interaction where each singularity can be of collinear or multi-particle type with codimension one. Therefore each diagram of this type has two different poles which can be completely determined by two variables. A soft singularity arises where an external momentum vanishes, and as a result the contribution of this process to the amplitude contains a singular factor. For a soft momentum of a massless particle, there are again two equations to determine the pole, since each index $\alpha$ or $\dot{\alpha}$ in $P_{\alpha\dot{\alpha}}$ runs over 1 and 2, hence we need exactly two variables to solve the equations. 

Having linear deformations, one can apply Cauchy's theorem to the complexified amplitude which has now two linear polynomials in the denominator, 
$$\oint\oint dz_1dz_2 \dfrac{1}{(az_1+bz_2+c)(a'z_1+b'z_2+c')}.$$
These polynomials are denominators of propagators, which become on-shell, or of the soft factors. We first carry out, e.g., $z_2$-integral in which the corresponding pole is considered as a function of the other variable, $z_2^*=z_2^*(z_1)$. We will finally find $1/(ab'-a'b)$ after the second integration. The same result can be obtained from the global residue theorem which will be discussed in the next section. 

With generic deformations, propagators will have higher degree polynomials in denominators corresponding to different types of singularities.  
In case these polynomials are irreducible, Cauchy's theorem does not work and the global residue theorem has to be applied. This theorem, the generalization of Cauchy's one variable residue theorem, is the only tool in calculations with more complex variables and higher degrees. 

Toward having the goal of presenting amplitudes which makes different singularities manifest, in this paper we extend Risager's deformation to $\mathbb{C}^2$ and will see amplitudes expose codimension two singularities. As was discussed in the introduction, Risager's shift naturally needs to be defined in a two-complex-dimensional plane. Therefore, our generalized deformation on negative helicities will be,
$$\tilde{\lambda}^i\longrightarrow\tilde{\lambda}^i+\alpha^i(z_1\tilde{\zeta}_1+z_2\tilde{\zeta}_2),$$
with $\tilde{\zeta}_1$ and $\tilde{\zeta}_2$ being two reference spinors, and $\alpha^i$ are determined in such a way that momentum conservation is preserved. Although with this linear deformation it is possible to recover the non-deformed amplitude using Cauchy's theorem, we will apply the global residue theorem in our calculations.

\section{Review of residues in multi-dimensional complex analysis}
Starting by a linear deformation in two complex variables, $z_1$ and $z_2$, on $\tilde{\lambda}^i$'s of negative helicity particles, generalization of Risager's deformation, the amplitude will be a rational function of both variables. In analogy with one-variable analysis, we study the following contour integral from which the physical amplitude, $M(0,0)$, can be obtained
\begin{equation}
\oint dz_1dz_2\,\dfrac{M(z_1,z_2)}{z_1z_2},
\label{2var}
\end{equation}
where the denominator of $M(z_1,z_2)$ factorizes into pieces coming from deformed propagators. Therefore, the full integrand of \eqref{2var} can be written as $\dfrac{g(z_1,z_2)}{f_1(z_1,z_2)f_2(z_1,z_2)}$, where $z_1$, $z_2$ and the factors of the denominator of $M$ could arbitrarily belong to $f_1$ or $f_2$. The functions $f_1$, $f_2$ and $g$ are polynomials, and $g$ is regular at zeros of the denominator. 

Now, let $\Gamma$ be the set of all the zeros of $f_1$ and $f_2$,
\begin{equation}
\Gamma =\lbrace P=(z_1^*,z_2^*)\vert f_1(z_1^*,z_2^*)=f_2(z_1^*,z_2^*)=0\rbrace.
\end{equation}
The \textit{Global Residue Theorem} for any $f_1$ and $f_2$, states that
\begin{equation}
\sum_{P\in\Gamma} \text{Res}\left(\dfrac{M(z_1,z_2)}{z_1z_2}\right)_{\!\!P} =0,
\label{GRT}
\end{equation}
when the degree condition
\begin{equation}
\text{deg}(g)<\text{deg}(f_1)+\text{deg}(f_2)-2
\label{degree}
\end{equation}
is satisfied.\footnote{With $n$ complex variables and therefore $n$ maps, $(f_1,\cdots,f_n): \mathbb{C}^n\rightarrow\mathbb{C}^n$, the degree condition generalizes to $\text{deg}(g)<\text{deg}(f_1)+\cdots+\text{deg}(f_n)-n$. This condition is analogous to having no pole at infinity in the usual BCFW or Risager's deformation with one complex variable. }

Since there are different ways to group factors of the denominator into $f_1$ and $f_2$, there exist different residue theorems for a given function $M(z_1,z_2)$. Each term in \eqref{GRT} is a contour integral for small $\epsilon$ as follows,
\begin{align}
\text{Res}\left(\dfrac{g(z_1,z_2)}{f_1(z_1,z_2)f_2(z_1,z_2)}\right)_{\!\!P} = &\dfrac{1}{(2\pi i)^2}\oint _{|f_1|=\epsilon , |f_2|=\epsilon } dz_1dz_2\, \dfrac{g(z_1,z_2)}{f_1(z_1,z_2)f_2(z_1,z_2)} \notag \\
= &\dfrac{g(z^*_1,z^*_2)}{(2\pi i)^2} \oint_{|u|=|v|=\epsilon} \dfrac{du}{u}\frac{dv}{v}\, \text{det}\!\left(\dfrac{\partial(f_1,f_2)}{\partial(z_1,z_2)}\right)^{\! -1} 
\end{align}
where in the last line we performed a change of variables, $u=f_1$ and $v=f_2$, so the corresponding Jacobian, evaluated at $P=(z_1^*,z_2^*)$, appears inside the integral. 

As can be seen above, the integration factorizes into two pieces, each on a $\mathbb{C}^1$-plane similar to one variable analysis. The full contour is therefore $S^1\times S^1\subset \mathbb{C}^2$ which unlike the one variable case does not fully enclose the pole. This is in fact one important difference between one and several complex integrals. Each of these integrals around the defined contour equals 1, therefore the residue is given by
\begin{equation}
\text{Res}\left(\dfrac{g(z_1,z_2)}{f_1(z_1,z_2)f_2(z_1,z_2)}\right)_{\!\!P}= g(z_1^*,z_2^*)\,\text{det}\!\left(\dfrac{\partial(f_1,f_2)}{\partial(z_1,z_2)}\right)^{\! -1} \!\!\!\!(z_1^*,z_2^*).
\label{res}
\end{equation}

While having a determinant, the ordering of arguments is important. We fix the orientation of contours in such a way that $f_1$ always comes before $f_2$ in the Jacobian. With the order reversed there will be a minus sign for the residue.

Now in case $f_1$ contains $z_1$ and $f_2$ contains $z_2$, one possible solution for $f_1=f_2=0$ would be $z_1^*=z_2^*=0$. It is obvious that $M(z_1,z_2)$ is not singular at $(0,0)$ since this corresponds to no deformation on the amplitude. Therefore, \eqref{res} gives the physical non-deformed amplitude,
\begin{equation*}
\text{Res}\left(\dfrac{M(z_1,z_2)}{z_1z_2}\right)\!\!(0,0)=M(0,0).
\end{equation*}  
This simply fixes our convention for the definition of $f_1$ and $f_2$. In order for \eqref{GRT} to contain $M(0,0)$ as one of the terms, $z_1$ and $z_2$ have to belong to different functions. Applying this convention, we will use \eqref{GRT} and \eqref{res} for calculations in the following sections provided the degree condition is satisfied. 

In the same way that Risager \cite{Risager} proved the $z^{-k}$ behavior of $\text{N}^{k-2}$MHV Yang-Mills amplitudes, $k$ being the number of negative helicities, we show that the degree condition \eqref{degree} is satisfied with our 2-variable deformation. The most dangerous Feynman diagrams are those with only cubic vertices. Performing this deformation on $\tilde{\lambda}$ of all negative helicity particles, one finds that $\text{deg}(g)=m$ when there are $m$ cubic vertices. The reason is that each cubic vertex depending on a deformed momentum is linear in $z_1$ and $z_2$. 

In the denominator of amplitudes we have contributions from $m-1$ propagators and $k$ polarization vectors. Each propagator linearly depends on complex variables. On the other hand, the $z_{1,2}$-dependence of polarization vectors depends on their helicities. For negative helicities we have $$\epsilon^{\mu (-)}(p)=\dfrac{\lambda^{\alpha}(p)\sigma^{\mu}\tilde{\lambda}^{\dot{\alpha}}(q)}{\sqrt{2}[pq]},$$ where $q$ is an auxiliary spinor. The deformation is on $\tilde{\lambda}(p)$, so each polarization vector with negative helicity contributes a $+1$ to the degree of denominator. Since we are working with $M(z_1,z_2)/z_1 z_2$, the total degree of denominator will be $\text{deg}(f_1)+\text{deg}(f_2)=(m-1)+k+2$. The degree condition then says $m<m+k+1-2$ or $1<k$ which is true. Therefore, the validity condition of the global residue theorem is satisfied for our 2-variable deformation on $n$-particle $\text{N}^{k-2}$MHV Yang-Mills tree amplitudes.

\section{5-Particle NMHV tree amplitude with 2-variable shifts}

The aim is to calculate Yang-Mills tree amplitudes using the global residue theorem. As a case in point, we consider the split-helicity NMHV 5-particle amplitude, $M(- - - + +)$, under the deformation
\begin{equation}
    \hat{\tilde{\lambda}}^i(z_1,z_2)=\tilde{\lambda}^i+\alpha ^i\tilde{\eta}, 
\end{equation}
where $i=1,2,3$ and we choose $\tilde{\eta}=z_1 \tilde{\lambda}^4+z_2 \tilde{\lambda}^5$. Using momentum conservation, a nontrivial solution for $\alpha^i$ is $\alpha^i=\langle j k\rangle$ where $i$, $j$ and $k$ cyclically take values of 1, 2 and 3.

Since we are working with color-ordered amplitudes, the $z$-dependent propagators are those with $\hat{P}^2_{12},\hat{P}^2_{23},\hat{P}^2_{34},\text{and }\hat{P}^2_{51}$ which together with $z_1$ and $z_2$ are the factors in $f_1$ and $f_2$. The diagram with particles 4 and 5 being on one sub-diagram does not contribute, since $P_{4,5}^2$ has no $z_{1,2}$-dependence. 

The simplest choice
\begin{equation}
f_1=z_1, \quad f_2=z_2 \hat{P}^2_{1,2}\hat{P}^2_{2,3}\hat{P}^2_{3,4}\hat{P}^2_{5,1},
\label{ris}
\end{equation} 
results in the 1-variable Risager's deformation since one of the complex variables, $z_1$, is zero throughout calculations. 

\begin{figure}
\centering
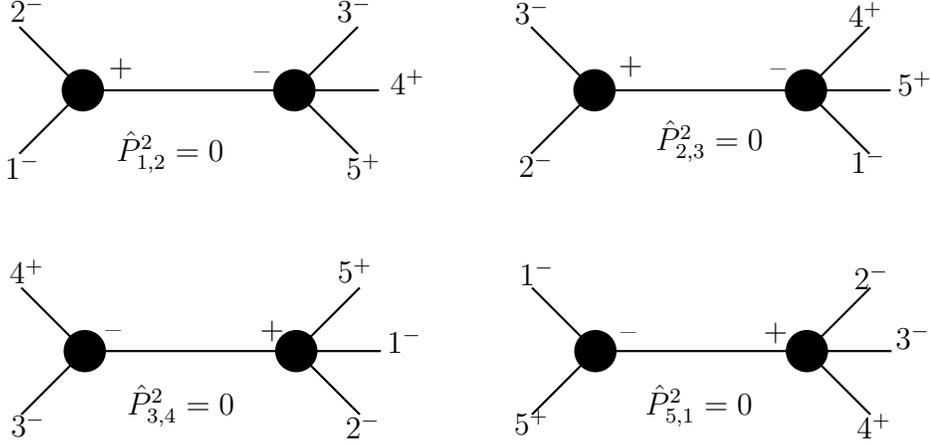
\caption{BCFW diagrams of 5-particle NMHV amplitude.}
\label{fig1}
\end{figure}

Apart from $(0,0)$, $f_1=f_2=0$ has 4 solutions. Hence, there are four terms, all with collinear/multi-particle singularities, in the sum of the residues,
\begin{equation}
      M(0,0)=-\sum_{poles\neq(0,0)} \text{Res}(\dfrac{M(z_1,z_2)}{z_1 z_2}),
\label{eq2}    
\end{equation}
corresponding to the four diagrams in Fig. \ref{fig1} \footnote{With this choice for $\tilde{\eta}$, $P^2_{1,5}$ is independent of both complex variables when $f_1=z_1=0$. Therefore, the residue corresponding to this channel vanishes.}.

In the next example we consider 
\begin{equation}
      f_1=z_1\hat{P}^2_{1,2}, \quad f_2=z_2\hat{P}^2_{2,3}\hat{P}^2_{3,4}\hat{P}^2_{5,1}.
\label{2exmpl}      
\end{equation}  
This time, solutions to $f_1=f_2=0$ are coming from 
\begin{equation}
     z_1=0, \left\{ \begin{array}{rcc}
          \hat{P}^2_{2,3}=0 \\
           \hat{P}^2_{3,4}=0 \\
           \hat{P}^2_{5,1}=0 
        \end{array} \right. \quad\quad \text{or} \quad\quad
     \hat{P}^2_{1,2}=0,  \left\{ \begin{array}{rcc}
           z_2=0\\
          \hat{P}^2_{2,3}=0 \\
           \hat{P}^2_{3,4}=0 \\
           \hat{P}^2_{5,1}=0 
        \end{array} ,\right.     
        \label{systeqs} 
\end{equation}
in addition to $z_1=z_2=0$ which corresponds to the non-deformed amplitude. 

Having $z_1=0$ or $z_2=0$ in any system of equations, the problem reduces to 1-variable Risager's deformation with a collinear singularity. The corresponding diagrams are exactly those in Fig. \ref{fig1}. 

\begin{figure}
\centering
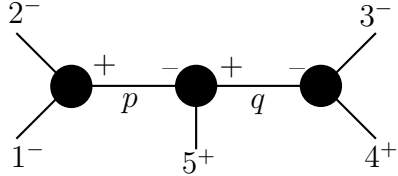
\caption{Double factorization channel in 5-particle NMHV amplitude.}
\label{fig2}
\end{figure}

Let's now consider $\hat{P}^2_{1,2}=\hat{P}^2_{3,4}=0$, Fig. \ref{fig2}, with solutions 
\begin{equation}
z_1^*=\frac{s_{12}-s_{34}}{\langle 12\rangle[54]\langle 35\rangle}, \quad z_2^*=\frac{[34]}{\langle12\rangle[45]}.
\label{sol1}
\end{equation}

In this double factorization channel we have three sub-amplitudes multiplying and forming the diagram,  
\begin{equation*}
      \frac{\langle 12\rangle^3}{\langle 2p\rangle\langle p1\rangle}\frac{1}{\langle 12\rangle[\hat{1}\hat{2}]}\frac{[q5]^3}{[5p]       [pq]}\frac{1}{\langle34\rangle[\hat{3}4]}\frac{\langle q3\rangle^3}{\langle 34\rangle\langle 4q\rangle}=\frac{[45]^3}{[1^*5][3^*2^*] [\hat{1}\hat{2}][\hat{3}4]},
\end{equation*} 
where the starred spinors are evaluated at \eqref{sol1} and the hatted ones depend on $z_1$ and $z_2$. Using \eqref{res}, the residue for this process can be obtained, 
\begin{equation}
     \frac{\langle12\rangle^3\langle35\rangle^2[45]}{\langle34\rangle\langle25\rangle\langle15\rangle\langle45\rangle[34](s_{12}-s_{34})}.
\end{equation}

The next system of equations is $\hat{P}^2_{1,2}=\hat{P}^2_{2,3}=0$ with a shared  deformed momentum $\hat p_2$. Similarly, in $\hat{P}^2_{1,2}=\hat{P}^2_{5,1}=0$, the last equations, $\hat p_1$ is shared. One can easily see that there is no way to draw a diagram with correct factorizations for any of the cases at hand. 
For a detailed examination of these processes we write the former as
\begin{equation}
    \left\{ \begin{array}{rc} 
    \left[\hat{1}\hat{2}\right]=0 \\   \left[\hat{2} \hat{3}\right]=0 \end{array} \right. ,      
\label{eq3}
\end{equation}
with some solutions $z^*_1$ and $z^*_2$. Simple calculations show that $\tilde{\lambda}^2(z^*_1,z^*_2)=0$. In fact from \eqref{eq3} one can see that evaluated at $(z_1^*,z_2^*)$, $\tilde{\lambda}^1 \Vert \tilde{\lambda}^2$ and $\tilde{\lambda}^2 \Vert \tilde{\lambda}^3$ but $\tilde{\lambda}^1$ and $\tilde{\lambda}^3$ are not parallel. Therefore we can conclude that $\tilde{\lambda}^2(z^*_1,z^*_2)=0$.

The shifted momentum of particle 2 is being soft. This means that the contribution of this channel comes from the soft limit $\hat{\tilde{\lambda}}^2\rightarrow0$ of the full amplitude.

In general, Yang-Mills tree amplitudes in the soft limit of one of the momenta factorize into two parts, an amplitude without the soft particle and a singular factor,
\begin{equation}
M_n(\ldots,i-1,i,i+1,\ldots)\xrightarrow {p_i\rightarrow 0} \text{Soft}(i-1,i,i+1) M_{n-1}(\ldots,i-1,i+1,\ldots),
\end{equation} 
where clearly $M_{n-1}$ has no singularity at the limit $p_i\rightarrow0$. 

The soft factor, first computed by Weinberg \cite{Weinberg}, in spinor-helicity notation is
\begin{equation}
\text{Soft}(i-1,i,i+1) = \begin{cases} \dfrac{\langle i\! -\! 1\, i\! +\! 1\rangle}{\langle i\! -\! 1\, i\rangle\langle i\, i\! +\! 1\rangle}, & \mbox{if } \lambda ^i\rightarrow 0 \\ \dfrac{[i\! -\! 1\, i\! +\! 1]}{[i\! -\! 1\, i][i\, i\! +\! 1]}, & \mbox{if } \tilde{\lambda}^i\rightarrow 0 \end{cases}.
\end{equation}
Having this behavior, one can find the residue of $\dfrac{M(z_1,z_2)}{z_1z_2}$ in this limit. 

For the case where $\tilde\lambda^2(z_1^*,z_2^*)$ is soft we plug $\dfrac{[\hat{1}\hat{3}]}{[\hat{1}\hat{2}][\hat{2}\hat{3}]}M(\hat1^-,\hat3^-,4^+,5^+)$ into \eqref{res}, and the residue will be
\begin{equation}
\dfrac{\langle13\rangle^3 [45]}{\langle34\rangle\langle45\rangle\langle51\rangle[42][52]}.
\end{equation} 
Similarly, the solutions of the system of equations $[5\hat{1}]=0$ and $[\hat{1}\hat{2}]=0$ satisfy $\tilde{\lambda}^1(z_1^*,z_2^*)=0$, and the contribution of this channel to the amplitude comes from
\begin{equation*}
M(\hat{1}^-,\hat{2}^-,\hat{3}^-,4^+,5^+)\xrightarrow {\hat{\tilde\lambda}^1\rightarrow 0} \dfrac{[5\hat{2}]}{[5\hat{1}][\hat{1}\hat{2}]}M(\hat2^-,\hat3^-,4^+,5^+),
\label{eq5}
\end{equation*} 
with the residue being
\begin{equation}
-\dfrac{\langle23\rangle^3[45]}{\langle34\rangle\langle45\rangle\langle52\rangle[41][51]}.
\end{equation}

Finally, we add up all the relevant terms and the known result of NMHV 5-particle amplitude can be obtained,
\begin{equation}
M(1^-,2^-,3^-,4^+,5^+)=\dfrac{[45]^3}{[12][23][34][51]}.
\end{equation}

One can consider other combinations in $f_1$ and $f_2$ and apply the residue theorem. In 5-particle amplitude for any choice of these functions there are always collinear (via single or double factorizations) and soft singularities.
  
In amplitudes with more particles we will have multi-particle singularities as well (in the 5-particle example collinear and multi-particle singularities are the same). This may result in some difficulties in finding the residues, as there will be more shared particles between simultaneous equations. We will see that these cases often result in vanishing residues, and soft singularities are the only ones in addition to previously known collinear and  multi-particle singularities. The double factorization channels, which also appear in the expansion, are in fact made out of collinear and/or multi-particle singularities.

\section{N$^{k-2}$MHV amplitudes}

For a general discussion on the singularities of 2-variable deformed amplitudes, we consider the most general N$^{k-2}$MHV amplitude where $k$ negative helicities are randomly distributed, $M_n(+,\cdots,i_1^-,\cdots,i_2^-,\cdots,i_k^-,\cdots,+)$. As before, the two variable deformations are only on negative helicities. Using $\tilde{\lambda}$'s of two particles, $\tilde{\eta}$ can be defined, and the deformation will be 
\begin{equation*}
\tilde{\lambda}^i\longrightarrow \tilde{\lambda}^i+ \alpha^i \tilde{\eta}(z_1,z_2).
\end{equation*}

There are infinite families of $\alpha^i$ for $k>3$ which can be turned into more complex variables. For $k=3$, as was seen in our 5-particle example, we can fix these coefficients $\alpha^a=\langle bc\rangle$ up to an overall factor, where $a$, $b$, and $c$ cyclically take the indices of negative helicities. 

Similar to previous example, we first determine the $z_{1,2}$-dependent propagators on which the factorizations take place. Hence, at least one but not all of the deformed momenta are included between particles $A$ and $B$ in the set of denominators of propagators, $\mathcal{P}=\{\hat{P}_{A,B}^2=(p_{_A}+\cdots+p_{_B})^2\}$. As stated before, $f_1(z_1,z_2)$ and $f_2(z_1,z_2)$ contain $z_1$ and $z_2$ respectively as well as an arbitrary grouping of elements of $\mathcal{P}$. We will explain what possible channels do contribute to the full amplitude by different ways of getting $f_1=f_2=0$.

In case $z_1=0$ or $z_2=0$, the one-variable shift, the corresponding residue follows from a collinear or multi-particle channel depending on how many particles are forming $\hat{P}_{A,B}^2$.  

For cases where two members of $\mathcal{P}$ simultaneously vanish,
\begin{equation}
    \left\{ \begin{array}{rc} 
    \hat{P}_{a,b}^2=0 \\   \hat{P}_{c,d}^2=0 \end{array} \right. ,      
\label{double}
\end{equation}
depending on how indices overlap, different events may happen. One can imagine various orderings and coincidences of particles as follows: $1) \  a<c<d<b$, $2) \  a<c<b=d$, $3) \ a<c<b<d$, $4) \ a<b=c<d$,
where any other distribution is equivalent to one of these cases. For instance, using momentum conservation one can see that the case where the two sets are completely separated, $a<b<c<d$, is exactly the first ordering which results in double factorization.  

\begin{figure}
\centering
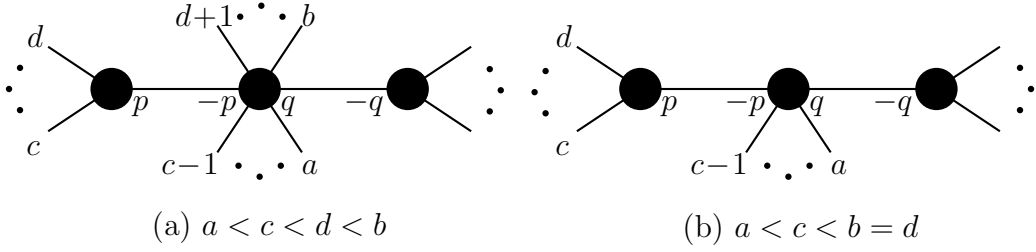
\caption{Double factorization channels in $\text{N}^{k-2}$MHV amplitude.}
\label{gen}
\end{figure}

Therefore, 1 says that the diagram has three sub-amplitudes, one with particles $\{c,\cdots, d,p\}$, the other in the middle with $\{a,\cdots,c -1,-p,d+1,\cdots,b,q\}$, and rest of the particles are in the third sub-amplitude as in Fig.\ref{gen}(a). The two singularities here can either be collinear or multi-particle.

Next, we have 2 again with double factorization, $\{a,\cdots,c-1,-p,q\}$ in the middle, $\{c,\cdots,d,p\}$ on the left and the rest in the third sub-amplitude, Fig.\ref{gen}(b). Again as in 1, the process can have two collinear or multi-particle singularities.

In 3, the overlap is again non-empty but we can not find any Feynman diagram associated with the given propagators. In fact the amplitude can not factorize in this way. One can also check that there is no soft singularity at the solutions of \eqref{double}, $(z_1^*,z_2^*)$. Therefore, $(z_1^*,z_2^*)$ does not correspond to any pole of the amplitude. We support our argument by explicit evaluation of the residue of $M_6(-+-+-+)$ at $(z_1^*,z_2^*)$ and find that it vanishes.

In 4, the two sets share only a single particle. This may lead us to conclude that the shared particle, if deformed, is soft as was seen in the 5-particle example. It is true only if both singularities are collinear. To see this, assume that $\tilde{\lambda}^b(z^*_1,z^*_2)\rightarrow0$ is a solution to \eqref{double} when $b=c$. Hence we will have $\hat{P}^2_{a,b-1}=0$ and $\hat{P}^2_{b+1,d}=0$ which are independent of $\tilde{\lambda}^b$ and therefore are not necessarily valid unless $a=b-1$ and $b+1=d$. Having said that, \eqref{double} reduces to $[a b]=[b d]=0$ which is equivalent to both singularities being collinear. 

One may also imagine a case where the two sets of indices coincide, $a=c<b=d$. In fact this can never happen to Feynman diagrams since there is no double pole in propagators.

We conclude that using 2-variable Risager's deformation in $n$-point N$^{k-2}$MHV amplitudes, collinear, multi-particle and soft singularities of tree amplitudes can be probed. With a generic one-variable shift, soft channels do not contribute to amplitudes and it is the second complex variable which is necessary for probing these channels.

\section{6-Particle NMHV amplitude}
In this section we compute the 6-particle amplitude with alternating helicities, $M(-+-+-+)$. We choose particles 2 and 4 in the definition of the reference spinor. Since each pair of adjacent momenta in this helicity configuration is deformed, the number of complex propagators with Risager's shift is maximum. Together with three multi-particle propagators, we can arbitrarily define $f_1$ and $f_2$, e.g., 
$$f_1(z_1,z_2)=z_1\hat{P}^2_{1,2}\hat{P}^2_{4,5}\hat{P}^2_{1,3}, \quad f_2(z_1,z_2)=z_2 \hat{P}^2_{2,3}\hat{P}^2_{3,4}\hat{P}^2_{5,6}\hat{P}^2_{6,1}\hat{P}^2_{2,4}\hat{P}^2_{3,5}.$$   

\begin{figure}
\centering
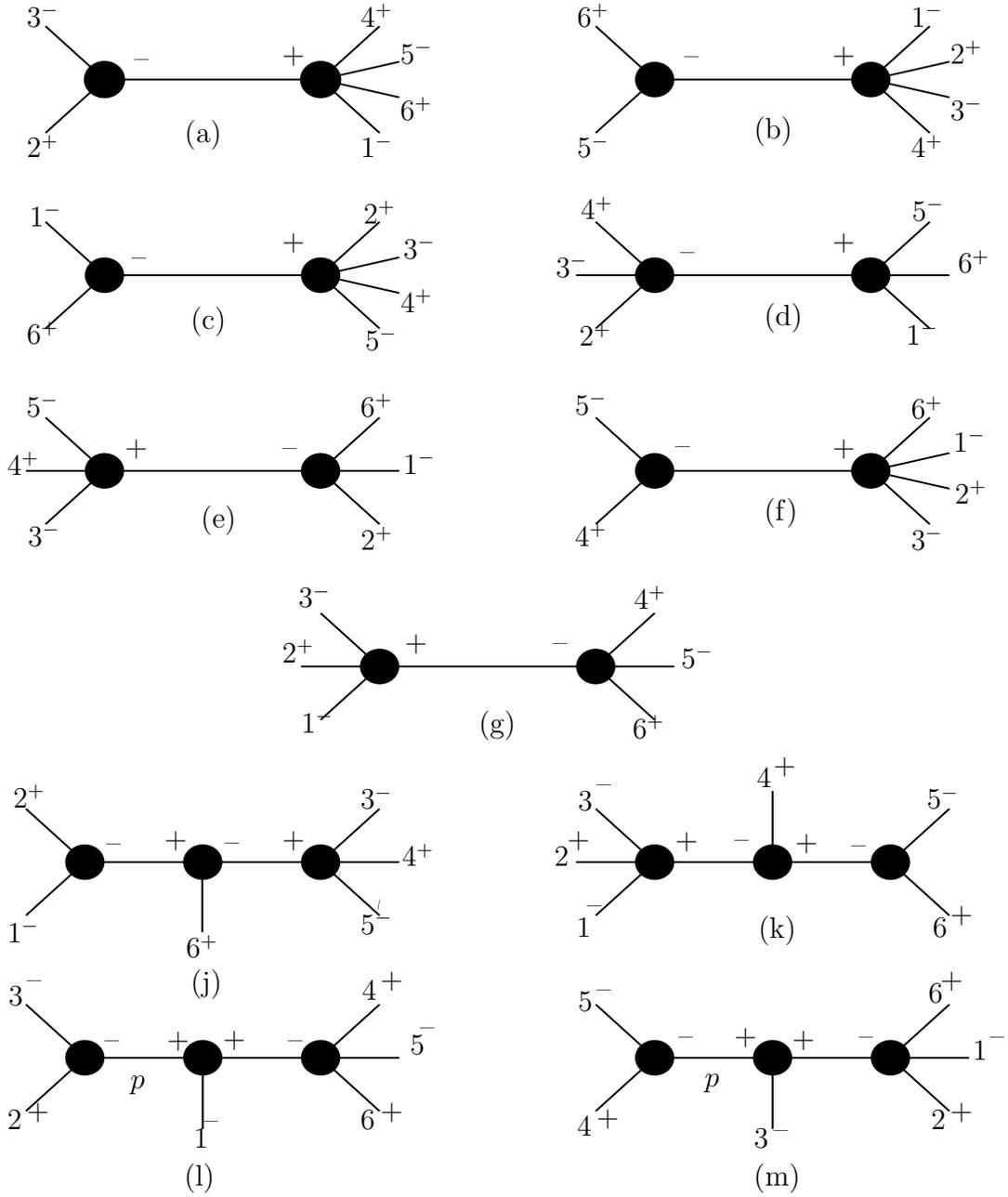
\caption{BCFW diagrams of 6-particle NMHV amplitude.}
\label{fig4}
\end{figure}

As was discussed before, the contributions from $z_1=0$ or $z_2=0$ are the usual Risager's terms. The corresponding residues are, Fig. \ref{fig4}, 
\begin{align}
\hspace{-70pt}\text{(a)}\ z_1=\hat{P}^2_{2,3}=0:\quad &\frac{\langle 15\rangle^4[24]^3}{\langle 45\rangle\langle 56\rangle\langle 61\rangle[23][34]\langle 1\vert 2+3\vert 4]\langle 4\vert 2+3\vert 4]}, 
\end{align}
\vspace{-20pt}
\begin{align}
\hspace{-70pt}\text{(b)}\ z_1=\hat{P}^2_{5,6}=0:\quad &\frac{\langle 13\rangle^4[64]^3}{\langle 12\rangle\langle 23\rangle\langle 34\rangle[45][56]\langle 1\vert 2+3\vert 4]\langle 4\vert 5+6\vert 4]},
\end{align}
\vspace{-20pt}
\begin{align}
\hspace{-70pt}\text{(c)}\ z_1=\hat{P}^2_{6,1}=0:\quad &\frac{\langle 35\rangle^4[46]^3}{\langle 23\rangle\langle 34\rangle\langle 45\rangle[61][14]\langle 5\vert 2+3\vert 4]\langle 2\vert 1+6\vert 4]},
\end{align}
\vspace{-20pt}
\begin{align}
\text{(d)}\ z_1=\hat{P}^2_{2,4}=0:\quad &\frac{-\langle 15\rangle^4\langle 23\rangle^2[24]^4}{\langle 34\rangle\langle 56\rangle\langle 61\rangle[34]P^2_{2,4} \langle 1\vert 2+3\vert 4]\langle 4\vert 2+3\vert 4]\langle 5\vert 2+3\vert4]},
\end{align}
\vspace{-20pt}
\begin{align}
\hspace{-30pt}\text{(e)}\ z_1=\hat{P}^2_{3,5}=0:\quad &\frac{\langle 35\rangle^2\langle1\vert 3+5\vert4]^4}{\langle12\rangle\langle 34\rangle\langle 45\rangle\langle 61\rangle[34][45]P^2_{3,5}\langle6\vert3+5\vert4]\langle2\vert1+6\vert4]} ,
\end{align}
\vspace{-20pt}
\begin{align}
\hspace{-70pt}\text{(f)}\ \hat{P}^2_{4,5}=z_2=0:\quad &\frac{-\langle 13\rangle^4[24]^3}{\langle 61\rangle\langle 12\rangle\langle 23\rangle[45][52]\langle 3\vert 4+5\vert 2]\langle 6\vert 4+5\vert 2]},
\end{align}
\vspace{-20pt}
\begin{align}
\hspace{-30pt}\text{(g)}\ \hat{P}^2_{1,3}=z_2=0:\quad &\frac{\langle 13\rangle^2\langle 5\vert 1+3\vert2]^4}{\langle12\rangle\langle 23\rangle\langle 45\rangle\langle 56\rangle[12][23]P^2_{1,3}\langle6\vert 4+5\vert2]\langle4\vert1+3\vert2]}.
\end{align}
There is no contribution from $z_1=\hat{P}^2_{3,4}=0$ and $z_2=\hat{P}^2_{1,2}=0$ since the particles we chose in the definition of $\tilde{\eta}$ make both $P^2_{3,4}$ and $P^2_{1,2}$ independent of $z_1$ and $z_2$. Therefore the corresponding residues vanish.

Soft channels appear where $\hat{P}^2_{1,2}=\hat{P}^2_{6,1}=0$, and $\hat{P}^2_{4,5}=\hat{P}^2_{5,6}=0$ with $\hat{p}_1$ and $\hat{p}_5$ being zero respectively.
\begin{align}
\text{(h)}\ \hat{P}^2_{1,2}=\hat{P}^2_{6,1}=0:\quad &\frac{\langle 35\rangle^4[42]}{\langle 23\rangle\langle 34\rangle\langle 45\rangle\langle 56\rangle\langle 62\rangle[12][14]},\\
\text{(i)}\ \hat{P}^2_{4,5}=\hat{P}^2_{5,6}=0:\quad &\frac{\langle 13\rangle^4[42]}{\langle 12\rangle\langle 23\rangle\langle 34\rangle\langle 46\rangle\langle 61\rangle[45][52]}.
\end{align}
There are also double factorization channels,
\begin{align}
\text{(j)}\ \hat{P}^2_{1,2}=\hat{P}^2_{3,5}=0:\quad &\frac{-\langle 35\rangle^4\langle 61\rangle^2[24]}{\langle 34\rangle\langle 45\rangle\langle 56\rangle\langle 63\rangle\langle 26\rangle\langle 12\rangle[12]\langle 6\vert3+5\vert4]},\\
\text{(k)}\ \hat{P}^2_{1,3}=\hat{P}^2_{5,6}=0:\quad &\frac{\langle 13\rangle^4\langle 45\rangle^3[24]}{\langle 12\rangle\langle 23\rangle\langle 34\rangle\langle 41\rangle\langle 56\rangle\langle 64\rangle\langle 4\vert 5+6\vert 4]\langle 4\vert 1+3\vert 2]}.
\end{align}

Considering the diagram of $\hat{P}^2_{2,3}=\hat{P}^2_{1,3}=0$, one might naively guess that this channel vanishes. In fact it does if $\tilde\lambda^p$ is parallel to $\hat{\tilde\lambda}^1$, which is not true. More careful observation shows that $\lambda^p$ evaluated at the corresponding poles of the diagram is parallel to $\lambda^1$, and therefore the process does contribute to the amplitude. The same happens to the channel $\hat{P}^2_{4,5}=\hat{P}^2_{3,5}=0$ where here $\lambda^p$ is proportional to $\lambda^3$. Hence, we have two more double factorization channels contributing to the amplitude with residues,
\begin{align}
\text{(l)}\ \hat{P}^2_{1,3}=\hat{P}^2_{2,3}=0: \quad &\frac{-\langle15\rangle^4\langle13\rangle^2[24]}{\langle12\rangle\langle23\rangle\langle45\rangle\langle56\rangle\langle61\rangle\langle14\rangle
[23]\langle1\vert2+3\vert4]}, \\
\text{(m)}\ \hat{P}^2_{4,5}=\hat{P}^2_{3,5}=0: \quad &\frac{\langle13\rangle^4\langle35\rangle^2[24]}{\langle12\rangle\langle23\rangle\langle34\rangle\langle45\rangle\langle36\rangle\langle61\rangle
[45]\langle3\vert4+5\vert2]}.
\end{align}

Finally the full amplitude is given by \eqref{eq2} which agrees with the known result of 6-particle NMHV amplitude \cite{BCF}: 
\begin{align}
M(1^-,2^+,3^-,4^+,5^-,6^+)&= \frac{\langle 13\rangle^4[46]^4}{\langle 12\rangle\langle23\rangle[45][56]P^2_{1,3}\langle3\vert1+2\vert6]\langle1\vert2+3\vert4]} \notag \\
&+ \frac{[26]^4\langle35\rangle^4}{[61][12]\langle34\rangle\langle45\rangle P^2_{3,5}\langle3\vert4+5\vert6]\langle5\vert4+3\vert2]} 
\label{6part} \\ &+\frac{\langle 15\rangle^4[24]^4}{\langle 23\rangle\langle34\rangle[56][61]P^2_{2,4}\langle5\vert4+3\vert2]\langle1\vert2+3\vert4]} \notag.
\end{align}

\section{Conclusion}
In this paper, we studied the analytic structure of Yang-Mills tree level scattering amplitudes by a new deformation on external momenta. Using the power of multi-variable complex analysis, especially the global residue theorem, physical amplitudes can be written recursively in a way similar to BCFW method. The degree condition, under which the global residue theorem is valid, was proved for generic $n$-particle $\text{N}^{k-2}$MHV amplitudes where the two-complex-variable deformation is on $\tilde\lambda$ of ($-$) helicities.

While with a generic one variable Risager's deformation, collinear and multi-particle singularities of tree amplitudes can be probed, the generalized 2-variable shift reveals soft channels as well as the two other types of singularity. This generalization is actually the natural way of defining all negative helicity deformation since it allows the deformed $\tilde{\lambda}$'s to live on the entire $\mathbb{C}^2$. We computed Yang-Mills 5- and 6-particle NMHV amplitudes at tree level with this generalization and discussed that in a general $n$-particle $\text{N}^{k-2}$MHV amplitude the only singularities that can exist correspond to collinear, multi-particle and soft channels.

For each collinear or multi-particle singularity, there is one condition on the sum of external momenta. This means that one complex variable is enough to solve the condition and find the corresponding pole, as was the case in BCFW and Risager's deformations. On the other hand, softness of particles results in two conditions. Having two equations, we need two complex variables in the deformation as was shown in our work. This simply indicates the necessity of introducing more variables in the deformations when there are more types of singularity to be investigated. 

\begin{figure}
\centering
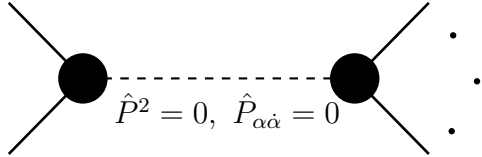
\caption{collinear-soft singularity}
\label{fig0}
\end{figure}

As an example of higher codimension singularities, one can consider a BCFW-like shift by which an internal deformed momentum is not only on-shell ($P^2=0$), but also soft ($P_{\alpha\dot{\alpha}}=\lambda_{\alpha}\tilde{\lambda}_{\dot{\alpha}}=0$). Here we have a codimension 3 singularity. Therefore, there are three conditions to define the poles for which we need three complex variables.  Figure \ref{fig0} shows a collinear-soft singularity of codimension 3.

There could be other deformations one can consider to investigate more interesting singularities. Under the experienced deformations, depending on which spinors are deformed, either $\langle ij\rangle$ or $[ij]$ is the singularity of an on-shell propagator with $P^2=\langle ij\rangle[ij]$. As an example, in \eqref{6part} we have both $\langle 12\rangle$ and $[12]$, but only the latter is singular under our two-dimensional Risager's deformation. A generalized deformation could present amplitudes in such a way that both of these brackets vanish. 

Another interesting example is double scalar soft limit in $\mathcal{N}=8$ SUGRA amplitudes at tree-level \cite{susy}. While the emission of a single soft scalar makes the amplitude vanish, the double soft limit does not commute. This is a reflection of the $E_{7(7)}$ global symmetry of the theory. This codimension 4 singularity could also be analyzed using BCFW-like deformations with more complex variables.

Probing new singularities can be thought of as part of the motivation for using multi-dimensional complex analysis. We also saw that the global residue theorem has to be used in generic cases while having higher degree singularities with more than one complex variable. 

\section*{Acknowledgements}
We are indebted to Freddy Cachazo for suggesting this problem and numerous enlightening discussions. We would like to thank Freddy Cachazo and Jorge Escobedo for useful comments on the draft of this paper. Research at Perimeter Institute is supported by the Government of Canada through Industry Canada and by the Province of Ontario through the Ministry of Research \& Innovation.


\begin{thebibliography}{99}


\bibitem{CSW}
  F.~Cachazo, P.~Svrcek, E.~Witten,
 ``MHV vertices and tree amplitudes in gauge theory,''
  JHEP {\bf 0409}, 006 (2004).
  [hep-th/0403047].


\bibitem{BCF}
  R.~Britto, F.~Cachazo, B.~Feng,
 ``New recursion relations for tree amplitudes of gluons,''
  Nucl.\ Phys.\  {\bf B715}, 499-522 (2005).
  [hep-th/0412308].

\bibitem{BCFW}
  R.~Britto, F.~Cachazo, B.~Feng {\it et al.},
``Direct proof of tree-level recursion relation in Yang-Mills theory,''
  Phys.\ Rev.\ Lett.\  {\bf 94}, 181602 (2005).
  [hep-th/0501052].


\bibitem{Bern review}
  Z.~Bern, L.~J.~Dixon, D.~A.~Kosower,
  ``On-Shell Methods in Perturbative QCD,''
  Annals Phys.\  {\bf 322}, 1587-1634 (2007).
  [arXiv:0704.2798 [hep-ph]].


\bibitem{susy1}
  A.~Brandhuber, P.~Heslop, G.~Travaglini,
  Phys.\ Rev.\  {\bf D78}, 125005 (2008).
  [arXiv:0807.4097 [hep-th]].


\bibitem{susy}
  N.~Arkani-Hamed, F.~Cachazo, J.~Kaplan,
  ``What is the Simplest Quantum Field Theory?,''
  JHEP {\bf 1009}, 016 (2010).
  [arXiv:0808.1446 [hep-th]].

\bibitem{duality}
  N.~Arkani-Hamed, F.~Cachazo, C.~Cheung {\it et al.},
 ``A Duality For The S Matrix,''
  JHEP {\bf 1003}, 020 (2010).
  [arXiv:0907.5418 [hep-th]].

\bibitem{G1}
  N.~Arkani-Hamed, J.~L.~Bourjaily, F.~Cachazo {\it et al.},
  ``The All-Loop Integrand For Scattering Amplitudes in Planar N=4 SYM,''
  JHEP {\bf 1101}, 041 (2011).
  [arXiv:1008.2958 [hep-th]].
  M.~Bullimore, L.~J.~Mason, D.~Skinner,
  ``MHV Diagrams in Momentum Twistor Space,''
  JHEP {\bf 1012}, 032 (2010).
  [arXiv:1009.1854 [hep-th]].
  N.~Arkani-Hamed, J.~Bourjaily, F.~Cachazo {\it et al.},
  ``Unification of Residues and Grassmannian Dualities,''
  JHEP {\bf 1101}, 049 (2011).
  [arXiv:0912.4912 [hep-th]].
  M.~Spradlin, A.~Volovich,
  ``From Twistor String Theory To Recursion Relations,''
  Phys.\ Rev.\  {\bf D80}, 085022 (2009).
  [arXiv:0909.0229 [hep-th]].
  L.~J.~Mason, D.~Skinner,
  ``Dual Superconformal Invariance, Momentum Twistors and Grassmannians,''
  JHEP {\bf 0911}, 045 (2009).
  [arXiv:0909.0250 [hep-th]].
  N.~Arkani-Hamed, F.~Cachazo, C.~Cheung,
  ``The Grassmannian Origin Of Dual Superconformal Invariance,''
  JHEP {\bf 1003}, 036 (2010).
  [arXiv:0909.0483 [hep-th]].
  M.~Bullimore, L.~J.~Mason, D.~Skinner,
  ``Twistor-Strings, Grassmannians and Leading Singularities,''
  JHEP {\bf 1003}, 070 (2010).
  [arXiv:0912.0539 [hep-th]].
  N.~Arkani-Hamed, J.~Bourjaily, F.~Cachazo {\it et al.},
  ``Local Spacetime Physics from the Grassmannian,''
  [arXiv:0912.3249 [hep-th]].
  J.~M.~Drummond, L.~Ferro,
  ``Yangians, Grassmannians and T-duality,''
  JHEP {\bf 1007}, 027 (2010).
  [arXiv:1001.3348 [hep-th]].
  J.~M.~Drummond, L.~Ferro,
  ``The Yangian origin of the Grassmannian integral,''
  JHEP {\bf 1012}, 010 (2010).
  [arXiv:1002.4622 [hep-th]].
  G.~P.~Korchemsky, E.~Sokatchev,
  ``Superconformal invariants for scattering amplitudes in N=4 SYM theory,''
  Nucl.\ Phys.\  {\bf B839}, 377-419 (2010).
  [arXiv:1002.4625 [hep-th]].

\bibitem{Risager}
  K.~Risager, "A direct proof of the CSW rules,"
  \textit{JHEP}\ {\bf 0512} (2005) 003 hep-th/0508206.

\bibitem{elvang2}
  H.~Elvang, D.~Z.~Freedman, M.~Kiermaier,
  ``Proof of the MHV vertex expansion for all tree amplitudes in N=4 SYM theory,''
  JHEP {\bf 0906}, 068 (2009).
  [arXiv:0811.3624 [hep-th]].


\bibitem{Weinberg}
  S.~Weinberg,
  ``Photons and Gravitons in s Matrix Theory: Derivation of Charge Conservation and Equality of Gravitational and Inertial Mass,''
  Phys.\ Rev.\  {\bf 135}, B1049-B1056 (1964).

\bibitem{CS}
  F.~Cachazo, P.~Svrcek,
 ``Tree level recursion relations in general relativity,''
  
  [hep-th/0502160].

\bibitem{witten}
  E.~Witten,
  ``Perturbative gauge theory as a string theory in twistor space,''
  Commun.\ Math.\ Phys.\  {\bf 252}, 189-258 (2004).
  [hep-th/0312171].


\bibitem{Nima Jared}
  N.~Arkani-Hamed, J.~Kaplan,
  ``On Tree Amplitudes in Gauge Theory and Gravity,''
  JHEP {\bf 0804}, 076 (2008).
  [arXiv:0801.2385 [hep-th]].

\bibitem{l11}
  F.~Cachazo, D.~Skinner,
  ``On the structure of scattering amplitudes in N=4 super Yang-Mills and N=8 supergravity,''
  
  [arXiv:0801.4574 [hep-th]].

\bibitem{taming}
  P.~Benincasa, C.~Boucher-Veronneau, F.~Cachazo,
  ``Taming Tree Amplitudes In General Relativity,''
  JHEP {\bf 0711}, 057 (2007).
  [hep-th/0702032 [HEP-TH]].


\bibitem{Parke-Taylor}
  S.~J.~Parke, T.~R.~Taylor,
  ``An Amplitude for n Gluon Scattering,''
  Phys.\ Rev.\ Lett.\  {\bf 56}, 2459 (1986).


\bibitem{GH}
P.~Griffiths and J.~Harris, "Principles of Algebraic Geometry," John Wiley \& Sons, 1994.

\bibitem{Dixon}
  L.~J.~Dixon,
  ``Calculating scattering amplitudes efficiently,''
  
  [hep-ph/9601359].


\bibitem{elvang1}
  H.~Elvang, D.~Z.~Freedman, M.~Kiermaier,
  ``Recursion Relations, Generating Functions, and Unitarity Sums in N=4 SYM Theory,''
  JHEP {\bf 0904}, 009 (2009).
  [arXiv:0808.1720 [hep-th]].


\end{thebibliography}
\end{document}